\documentclass[11pt]{article}
\usepackage{amssymb,amsmath,amsfonts}
\usepackage{graphicx}
\usepackage{graphics}
\usepackage{eepic,epsfig}

\begin{document}

\title{Electromagnetic Casimir effect for conducting plates \\
in de Sitter spacetime}
\author{A. S. Kotanjyan\thanks{%
E-mail: anna.kotanjyan@ysu.am}, A. A. Saharian\thanks{%
E-mail: saharian@ysu.am}, H. A. Nersisyan \\
\\
\textit{Department of Physics, Yerevan State University,}\\
\textit{1 Alex Manoogian Street, 0025 Yerevan, Armenia}}
\maketitle

\begin{abstract}
Two-point functions, the mean field squared and the vacuum expectation value
(VEV) of the energy-momentum tensor are investigated for the electromagnetic
field in the geometry of parallel plates on background of $(D+1)$%
-dimensional dS spacetime. We assume that the field is prepared in the
Bunch-Davies vacuum state and on the plates a boundary condition is imposed
that is a generalization of the perfectly conducting boundary condition for
an arbitrary number of spatial dimensions. It is shown that for $D\geq 4$
the background gravitational field essentially changes the behavior of the
VEVs at separations between the plates larger than the curvature radius of
dS spacetime. At large separations, the Casimir forces are proportional to
the inverse fourth power of the distance for all values of spatial dimension
$D\geq 3$. For $D\geq 4$ this behavior is in sharp contrast with the case of
plates in Minkowski bulk where the force decays as the inverse $(D+1)$th
power of the distance.
\end{abstract}

\bigskip


\bigskip

\section{Introduction}

The nontrivial properties of the quantum vacuum are manifested in its
response to external influences. In the present paper, for the
electromagnetic vacuum, we consider a problem with two kinds of influences:
background gravitational field and boundaries constraining the zero-point
fluctuations of the electromagnetic field. In order to have an exactly
solvable problem we study highly symmetric bulk and boundary geometries.
Namely, we consider the de Sitter (dS) spacetime as a background geometry
and two parallel plates with perfectly conducting boundary conditions. By
taking into account that the higher-dimensional models play an important
role in high-energy physics, in particular, in string theories and in
supergravity, we shall assume a general value of the spatial dimension $D$.
The dS spacetime is among the most popular backgrounds in the gravitational
physics. This geometry is of key importance in the inflationary cosmology
\cite{Lind}, according to which at the early stages of the cosmological
expansion our Universe passed through a phase with the geometry well
approximated by dS spacetime. Quantum fluctuations during an inflationary
epoch generate inhomogeneities which play a central role in the formation of
large scale cosmic structures. From the other side, at the present epoch the
Universe is accelerating and the expansion can be well approximated by a
homogeneous and isotropic model with a positive cosmological constant as the
dark energy. In this case, the dS spacetime is a future attractor for the
large scale geometry of the Universe. Consequently, the investigation of
physical effects in dS background is important for understanding the both
early and late stages of the Universe expansion.

In quantum field theory, the interaction of fluctuating quantum fields with
a classical background gravitational field leads to the polarization of the
vacuum. For the dS geometry this effect has been widely discussed in the
literature (for reviews see \cite{Birr82}). The presence of boundaries with
prescribed boundary conditions on the field operator gives rise to another
type of vacuum polarization known as the Casimir effect \cite{Most97}. The
boundaries can have different physical natures like macroscopic bodies in
QED, interfaces separating different phases, horizons in gravitational
physics, the branes in higher-dimensional field-theoretical models and in
string theories. Historically, the first example of the Casimir effect has
been considered for the electromagnetic field in the geometry of two neutral
parallel metallic plates \cite{Casi48}. Casimir showed that the plates
attract each other through the force which is proportional to the inverse
fourth power of the distance between them. The vacuum energy density and the
stresses are located in the region between the plates and their distribution
in this region is uniform \cite{Brow69}. The latter property is a
consequence of the conformal invariance of the electromagnetic field in
4-dimensional spacetime. The vacuum expectation values (VEVs) of the mean
electric and magnetic field squared have been discussed in \cite%
{Witt75,Lutk85}. These VEVs are not uniform and they diverge on the plates.
The Casimir densities for the electromagnetic field in an arbitrary number
of spatial dimensions $D$ are investigated in \cite{Alne06}. For $D>3$, the
electromagnetic field is not conformally invariant and the distributions of
the vacuum energy density and of the stresses parallel to the plates are not
uniform (see below). In all these considerations of the electromagnetic
Casimir effect for parallel conducting plates the background geometry was
Minkowskian. The Casimir force with the same geometry of boundaries and in
the presence of an extra dimension compactified on a circle is investigated
in \cite{Eder08}.

An important direction in the investigations of the Casimir effect is the
explicit dependence of the characteristics of the vacuum state on the
geometry of the background spacetime. The scalar Casimir densities in dS
spacetime, induced by planar and spherical boundaries with Robin boundary
conditions, are studied in \cite{Saha09} (special cases of conformally and
minimally coupled massless scalar fields are considered in \cite{Seta01}).
It has been shown that the background gravitational field decisively
influences the VEVs of physical observables at distances from the boundaries
larger than the curvature scale of the dS spacetime. For the same
background, similar features were observed in the topological Casimir
effect, induced by toroidal compactification of spatial dimensions \cite%
{Saha08}. The scalar vacuum densities in background spacetimes with dS and
anti-de Sitter (AdS) bubbles are considered in \cite{Bell14}. The
electromagnetic Casimir densities for a conducting plate in dS spacetime
have been investigated in \cite{Saha14}. The case of the electromagnetic
field in background of Friedmann-Robertson-Walker cosmological models with
power-law scale factors is discussed in \cite{Bell13} (for the
electromagnetic Casimir effect in Randall-Sundrum braneworld models on AdS
bulk see \cite{Teo10} and references therein).

The present paper is organized as follows. In the next section, by using the
complete set of the mode functions for the electromagnetic field, we
construct the two-point function for the electromagnetic field tensor in the
geometry of two conducting plates in background of $(D+1)$-dimensional dS
spacetime. This function is presented as the sum of boundary-free two-point
functions corresponding to an infinite sequence of image sources. The mean
electric field squared and the VEV of the energy-momentum tensor in the
region between two plates are investigated in Section \ref{sec:CasDens}.
Various limiting cases are discussed in detail. The Casimir forces acting on
the plates are studied in Section \ref{sec:Force}. The main conclusions of
the paper are summarized in Section \ref{sec:Conc}.

\section{Electromagnetic two-point functions}

\label{sec:TwoPoint}

Consider two perfectly conducting plates in the background of $\left(
D+1\right) $-dimensional dS spacetime. In inflationary coordinates, the
geometry is described by the line element%
\begin{equation}
ds^{2}=dt^{2}-e^{2t/\alpha }(d\mathbf{z})^{2},\;\mathbf{z}%
=(z^{1},z^{2},\ldots ,z^{D}),  \label{eq1}
\end{equation}%
where the parameter $\alpha $ is expressed in terms of the positive
cosmological constant $\Lambda $ by the relation $\alpha
^{2}=D(D-1)/(2\Lambda )$. Introducing the conformal time $\tau $ in
accordance with $\tau =-\alpha e^{-t/\alpha }$, $-\infty <\tau <0$, the
metric tensor is written in the conformally flat form: $g_{\mu \nu }=\left(
\alpha /\tau \right) ^{2}\eta _{\mu \nu }$ with $\eta _{\mu \nu }$ being the
metric tensor for Minkowski spacetime. In what follows we shall work in
spacetime coordinates $(\tau ,\mathbf{z})$.

We assume that the plates are placed at $z^{D}=0$ and $z^{D}=L$ and the
electromagnetic field is prepared in the Bunch-Davies vacuum state $%
|0\rangle $. Among the set of dS-invariant quantum states, the Bunch-Davies
vacuum is the only one for which the ultraviolet behavior of the two-point
functions is the same as in Minkowski spacetime. On the plates the field
obeys the boundary condition \cite{Ambj83} $n^{\nu _{1}}\,^{\ast }F_{\nu
_{1}\cdots \nu _{D-1}}=0$, where $^{\ast }F_{\nu _{1}\cdots \nu _{D-1}}$ is
dual to the electromagnetic field tensor $F_{\mu \nu }=\partial _{\mu
}A_{\nu }-\partial _{\nu }A_{\mu }$ and $n^{\mu }$ is the normal to the
plates. This condition is a generalization of the perfectly conducting
boundary condition in electrodynamics for models with an arbitrary number of
spatial dimensions. For the coordinates and for the momentum components
parallel to the plates we shall use the notations $\mathbf{z}_{\Vert
}=(z^{1},...,z^{D-1})$ and $\mathbf{k}_{\Vert }=(k_{1},...,k_{D-1})$. In the
regions $z^{D}<0$ and $z^{D}>L$ the VEVs are the same as those in the
geometry of a single plate located at $z^{D}=0$ and $z^{D}=L$, respectively.
This geometry is considered in \cite{Saha14} and below we shall be mainly
concerned with the region between the plates, $0<z^{D}<L$. In this region,
the mode functions for the vector potential $A_{\mu }(x)$, $x=(\tau ,\mathbf{%
z})$, realizing the Bunch-Davies vacuum state and obeying the gauge
conditions $A_{0}=0$, $\nabla _{\mu }A^{\mu }=0$, have the form%
\begin{equation}
A_{(\sigma \mathbf{k})l}(x)=iC\epsilon _{(\sigma )l}\eta
^{D/2-1}H_{D/2-1}^{(1)}(k\eta )\sin \left( k_{D}z^{D}\right) e^{i\mathbf{k}%
_{\parallel }\cdot \mathbf{z}_{\parallel }},  \label{Modes1}
\end{equation}%
for the components with $l=1,\ldots ,D-1$, and
\begin{equation}
A_{(\sigma \mathbf{k})D}(x)=C\epsilon _{(\sigma )D}\eta
^{D/2-1}H_{D/2-1}^{(1)}(k\eta )\cos \left( k_{D}z^{D}\right) e^{i\mathbf{k}%
_{\parallel }\cdot \mathbf{z}_{\parallel }},  \label{Modes2}
\end{equation}%
where $\eta =\left\vert \tau \right\vert $, $\mathbf{k}=(\mathbf{k}%
_{\parallel },k_{D})$, $k=\sqrt{k_{D}^{2}+\mathbf{k}_{\parallel }^{2}}$ and $%
\mathbf{k}_{\parallel }\cdot \mathbf{z}_{\parallel }\mathbf{=}%
\sum_{l=1}^{D-1}k_{l}z^{l}$. In these expressions, $H_{D/2-1}^{(1)}(k\eta )$
is the Hankel function of the first kind and for the polarization vectors $%
\epsilon _{(\sigma )l}$, with $\sigma =1,...,D-1$, one has the transverse
condition $\sum_{l=1}^{D}\epsilon _{(\sigma )l}k_{l}=0$ and the relations
\begin{equation}
\sum_{l=1}^{D}\epsilon _{(\sigma )l}\epsilon _{(\sigma ^{\prime })l}=\delta
_{\sigma \sigma ^{\prime }},\;\sum_{\sigma =1}^{D-1}\epsilon _{(\sigma
)l}\epsilon _{(\sigma )m}=\delta _{lm}-\frac{k_{l}k_{m}}{k^{2}}.
\label{PolVec}
\end{equation}

The modes (\ref{Modes1}) and (\ref{Modes2}) obey the boundary condition on
the plate $z^{D}=0$. From the boundary condition at $z^{D}=L$ for the
eigenvalues of the component of the momentum normal to the plates we get
\begin{equation}
k_{D}=\pi n/L,\text{ }n=0,1,...  \label{kD}
\end{equation}%
For the normalization coefficient one finds
\begin{equation}
|C|^{2}=\frac{(2\pi \alpha )^{3-D}}{2(1+\delta _{n0})L}.  \label{C2}
\end{equation}%
Note that for a general dS-invariant vacuum state, in the mode functions (%
\ref{Modes1}) and (\ref{Modes2}) a linear combination of the Hankel
functions $H_{D/2-1}^{(1)}(k\eta )$ and $H_{D/2-1}^{(2)}(k\eta )$ appears
instead of the function $H_{D/2-1}^{(1)}(k\eta )$. The choice of the
coefficient in the linear combination fixes the vacuum state.

Having a complete set of mode functions, we can evaluate the two-point
function $W_{lm}(x,x^{\prime })=\langle 0|A_{l}(x)A_{m}(x^{\prime
})|0\rangle $ for the vector potential by using the mode-sum formula
\begin{equation}
W_{lm}(x,x^{\prime })=\sum_{\sigma =1}^{D-1}\sum_{n=0}^{\infty }\int d%
\mathbf{k}_{\parallel }A_{(\sigma \mathbf{k})l}(x)A_{(\sigma \mathbf{k}%
)m}(x^{\prime }).  \label{Almmode}
\end{equation}%
By applying to the sum over $n$ the Poisson resummation formula, this
function is presented in the form
\begin{equation}
W_{lm}(x,x^{\prime })=\sum_{n=-\infty }^{\infty }\sum_{j=-1}^{+1}j^{1-\delta
_{mD}}W_{lm}^{(0)}(x,x_{j,n}^{\prime }),  \label{Wlm}
\end{equation}%
where $W_{lm}^{(0)}(x,x^{\prime })$ is the corresponding function in
boundary-free dS spacetime and%
\begin{equation}
x_{\pm 1,n}^{\prime }=\left( \tau ^{\prime },\mathbf{z}_{\parallel }^{\prime
},\pm z^{D\prime }+2nL\right) .  \label{xImage}
\end{equation}%
In (\ref{Wlm}), the term $j=+1$, $n=0$ corresponds to the boundary-free
function. The terms $j=-1$, $n=0$ and $j=-1$, $n=+1$ correspond to the parts
in the two-point function induced by single plates at $z^{D}=0$ and $z^{D}=L$%
, respectively, when the second plate is absent. The two-point functions for
both massive and massless vector fields in dS spacetime are considered in
\cite{Alle86}. Recently, it has been shown that the infrared pathologies of
the photon two-point function in dS spacetime are purely gauge artifacts
\cite{Yous11}.

A formula similar to (\ref{Wlm}) is obtained for the two-point function $%
W_{lm,pq}(x,x^{\prime })=\langle 0|F_{lm}(x)F_{pq}(x^{\prime })|0\rangle $
corresponding to the field tensor:%
\begin{equation}
W_{lm,pq}(x,x^{\prime })=\sum_{n=-\infty }^{\infty
}\sum_{j=-1}^{+1}j^{1-\delta _{pD}}W_{lm,pq}^{(0)}(x,x_{j,n}^{\prime }),
\label{WF}
\end{equation}%
with $W_{lm,pq}^{(0)}(x,x^{\prime })$ being the function for the
boundary-free dS geometry. Introducing the notations
\begin{equation}
B_{D}=\left( 4\pi \right) ^{\left( D-1\right) /2}\Gamma \left( \frac{D+3}{2}%
\right) ,\;z=1+\frac{(\eta -\eta ^{\prime })^{2}-\left\vert \Delta \mathbf{z}%
\right\vert ^{2}}{4\eta \eta ^{\prime }},  \label{BD}
\end{equation}%
\qquad with $\Delta \mathbf{z=z-z}^{\prime }$, for the components of the
boundary-free two-point function one has \cite{Saha14}
\begin{eqnarray}
W_{0l,0m}^{(0)}(x,x^{\prime }) &=&\frac{(\eta \eta ^{\prime })^{-2}}{%
2B_{D}\alpha ^{D-3}}\left[ \left( \delta _{lp}\delta _{mq}-\delta
_{lm}\delta _{pq}\right) \frac{\Delta z^{p}\Delta z^{q}}{2\eta \eta ^{\prime
}}\partial _{z}+(D-1)\delta _{lm}\right] G_{D}(z),  \notag \\
W_{lm,0p}^{(0)}(x,x^{\prime }) &=&\frac{(\eta \eta ^{\prime })^{-2}}{%
B_{D}\alpha ^{D-3}}\delta _{\lbrack lp}\delta _{m]q}\frac{\Delta z^{q}}{\eta
^{\prime }}\left[ 2+(z-\frac{\eta +\eta ^{\prime }}{2\eta })\partial _{z}%
\right] F_{D}(z),  \label{FF0} \\
W_{lm,pq}^{(0)}(x,x^{\prime }) &=&\frac{(\eta \eta ^{\prime })^{-2}}{%
B_{D}\alpha ^{D-3}}\left( \delta _{\lbrack lr}\delta _{m][p}\delta _{q]s}%
\frac{\Delta z^{r}\Delta z^{s}}{\eta \eta ^{\prime }}\partial _{z}+2\delta
_{\lbrack lp}\delta _{m]q}\right) F_{D}(z),  \notag
\end{eqnarray}%
where $l,m,p,q=1,2,\ldots ,D$, and the expression for the component $%
W_{0p,lm}^{(0)}(x,x^{\prime })$ is obtained from that for $%
W_{lm,0p}^{(0)}(x,x^{\prime })$ by changing the sign and by the interchange $%
\eta \rightleftarrows \eta ^{\prime }$. In (\ref{FF0}), the square brackets
enclosing the indices mean the antisymmetrization over these indices and we
have defined the functions%
\begin{eqnarray}
F_{D}\left( z\right) &=&\Gamma \left( D\right) F\left( D,2;\frac{D+3}{2}%
;z\right) ,  \notag \\
G_{D}\left( z\right) &=&2\Gamma \left( D-1\right) F\left( D-1,3;\frac{D+3}{2}%
;z\right) ,  \label{FD}
\end{eqnarray}%
with $F(a,b,c;z)$ being the hypergeometric function.

In what follows, we shall need the asymptotic expressions of the functions (%
\ref{FD}) (see also \cite{Saha14}). For $0<1-z\ll 1$, by using the
asymptotic expansion for the hypergeometric function \cite{Abra72}, to the
leading order one has%
\begin{equation}
F_{D}\left( z\right) \approx G_{D}\left( z\right) \approx (D+1)\frac{\Gamma
^{2}((D+1)/2)}{2(1-z)^{(D+1)/2}}.  \label{Fdas}
\end{equation}%
In the region $z\ll -1$, by making use of the linear transformation formulas
relating the hypergeometric functions with the arguments $z$ and $1/z$ \cite%
{Abra72}, for the function $F_{D}\left( z\right) $ we get%
\begin{eqnarray}
F_{D}\left( z\right) &\approx &(D^{2}-1)\frac{\Gamma (D-2)}{4(-z)^{2}},\;D>2,
\notag \\
F_{D}\left( z\right) &\approx &\frac{\Gamma ((D+3)/2)}{2^{D-1}\sqrt{\pi }%
(-z)^{D}}\Gamma \left( D\right) \Gamma (1-D/2),\;D<2.  \label{Fdas1}
\end{eqnarray}%
For $D=2$ one has $F_{D}\left( z\right) \approx 3(-z)^{-2}[\ln (-4z)-1]/4$.
In a similar way, for the function $G_{D}\left( z\right) $ one finds%
\begin{eqnarray}
G_{D}(z) &\approx &\frac{(D^{2}-1)(D-3)}{4(-z)^{3}}\Gamma (D-4),\;D>4,
\notag \\
G_{D}(z) &\approx &\Gamma (D-1)\frac{\Gamma ((D+3)/2)\Gamma (2-D/2)}{2^{D-3}%
\sqrt{\pi }(-z)^{D-1}},\;D<4,  \label{Gdas}
\end{eqnarray}%
and $G_{4}(z)\approx 3(-z)^{-3}\left[ 2\ln \left( -4z\right) -3\right] $ for
$D=4$.

On the base of the representation (\ref{WF}), the two-point function can be
presented in the decomposed form%
\begin{equation}
W_{lm,pq}(x,x^{\prime })=W_{lm,pq}^{(0)}(x,x^{\prime
})+W_{lm,pq}^{(b)}(x,x^{\prime }),  \label{Wdec}
\end{equation}%
where the boundary-induced part $W_{lm,pq}^{(b)}(x,x^{\prime })$ is given by
the right-hand side of (\ref{WF}) omitting the term with $j=+1$, $n=0$. Here
we consider a free field theory and all the properties of the vacuum state
are encoded in the two-point functions. In the next sections we consider the
Casimir densities and the forces acting on the plates.

\section{Casimir densities}

\label{sec:CasDens}

\subsection{Mean field squared}

We start the investigation of the VEVs in the region between the plates by
evaluating the mean electric field squared. For $D=3$ it determines the
Casimir-Polder force acting on a polarizable particle placed near the
plates. The VEV of the electric field squared is obtained from the two-point
function in the coincidence limit of the arguments:%
\begin{equation}
\left\langle E^{2}\right\rangle =-g^{00}g^{lm}\lim_{x^{\prime }\rightarrow
x}W_{0l,0m}(x,x^{\prime }).  \label{E2l}
\end{equation}%
For points away from the plates, the divergences in this limit are contained
in the boundary-free part and the renormalization is needed for this part
only. The renormalized VEV is presented in the form%
\begin{equation}
\left\langle E^{2}\right\rangle =\left\langle E^{2}\right\rangle
_{0}+\left\langle E^{2}\right\rangle _{\mathrm{b}},  \label{E2dec}
\end{equation}%
where $\left\langle E^{2}\right\rangle _{0}$ is the renormalized VEV in
boundary-free dS spacetime and the part $\left\langle E^{2}\right\rangle _{%
\mathrm{b}}$ is induced by the plates. The latter is obtained by the formula
similar to (\ref{E2l}) with the function $W_{0l,0m}^{(b)}(x,x^{\prime })$.

Introducing the notations
\begin{eqnarray}
u_{n} &=&1-\left( nL/\eta \right) ^{2},  \notag \\
v_{n} &=&1-\left( z^{D}-nL\right) ^{2}/\eta ^{2},  \label{vn}
\end{eqnarray}%
in the region between the plates the boundary-induced contribution is
presented in the form
\begin{eqnarray}
\left\langle E^{2}\right\rangle _{\mathrm{b}} &=&\frac{D-1}{2B_{D}\alpha
^{D+1}}\left\{ 2\sum_{n=1}^{+\infty }\left[ 2\left( u_{n}-1\right) \partial
_{u_{n}}+D\right] G_{D}(u_{n})\right.  \notag \\
&&\left. -\sum_{n=-\infty }^{+\infty }\left[ 2\left( v_{n}-1\right) \partial
_{v_{n}}+D-2\right] G_{D}(v_{n})\right\} .  \label{E2}
\end{eqnarray}%
This contribution depends on $z^{D}$, $L$ and $\eta $ through the
combinations $z^{D}/\eta $ and $L/\eta $. These ratios are the proper
distance from the left plate and the proper separation between the plates
measured in units of the dS curvature scale $\alpha $. This property is a
consequence of the maximal symmetry of the background spacetime and of the
Bunch-Davies vacuum state. In (\ref{E2}), the $n=0$ and $n=1$ terms of the
last series are the contributions induced by single plates at $z^{D}=0$ and $%
z^{D}=L$, respectively, when the second plate is absent. The surface
divergences are contained in these two terms and, hence, for points near the
plates they dominate in the VEV of the electric field squared. In
particular, near the left plate, $z^{D}/\eta \ll 1$, by using (\ref{Fdas}),
to the leading order one has%
\begin{equation}
\langle E^{2}\rangle _{\mathrm{b}}\approx \frac{3\left( D-1\right) \Gamma
((D+1)/2)}{2(4\pi )^{(D-1)/2}(\alpha z^{D}/\eta )^{D+1}}.  \label{E2bAs}
\end{equation}%
As is seen, near the plates the mean field squared is positive.

For $D=3$, by using $G_{3}(z)=2/(1-z)^{2}$, from the general formula one
finds
\begin{equation}
\left\langle E^{2}\right\rangle _{\mathrm{b}}=\frac{3\zeta
_{4}(z^{D}/L)-2\zeta (4)}{4\pi \left( \alpha L/\eta \right) ^{4}},
\label{E2D3}
\end{equation}%
with $\bigskip \zeta \left( s\right) $ being the Riemann zeta function, $%
\zeta (4)=\pi ^{4}/90$, and we have introduced the function%
\begin{equation}
\zeta _{p}(x)=\sum_{n=-\infty }^{\infty }\frac{1}{|n-x|^{p}}=\zeta
(p,x)+\zeta (p,1-x).  \label{Zetap}
\end{equation}%
Here $\zeta (p,x)$ is the Hurwitz zeta function. The electromagnetic field
is conformally invariant for $D=3$ and the expression (\ref{E2D3}) is
related to the corresponding result for the plates in Minkowski spacetime
with the separation $L$ by the formula $\left\langle E^{2}\right\rangle _{%
\mathrm{b}}=(\eta /\alpha )^{4}\left\langle E^{2}\right\rangle _{\mathrm{b}%
}^{(M)}$. In the case $D=5$, by making use of the expression $%
G_{5}(z)=12/(1-z)^{3}$, we get%
\begin{equation}
\left\langle E^{2}\right\rangle _{\mathrm{b}}=\frac{3\zeta
_{6}(z^{D}/L)-2\zeta (6)}{4\pi ^{2}\left( \alpha L/\eta \right) ^{6}},
\label{E2bD5}
\end{equation}%
with $\zeta (6)=\pi ^{6}/945$.

Let us consider the mean field squared in the limits of small and large
proper separations between the plates. When the proper separation of the
plates is small compared to the dS curvature radius, $L/\eta \ll 1$, the
curvature effects are small, and one can see that, to the leading order, the
VEV of the electric field squared coincides with that for plates in
Minkowski spacetime with the separation $\alpha L/\eta $:
\begin{equation}
\left\langle E^{2}\right\rangle _{\mathrm{b}}\approx \frac{\left( D-1\right)
\Gamma \left( \left( D+1\right) /2\right) }{2\left( 4\pi \right) ^{\left(
D-1\right) /2}(\alpha L/\eta )^{D+1}}\left[ 3\zeta _{D+1}(z^{D}/L)-2\zeta
\left( D+1\right) \right] .  \label{E2close}
\end{equation}%
It can be seen that the expression in the right-hand side of (\ref{E2close})
is obtained from the results of \cite{Alne06} for $\left\langle
E_{D}^{2}\right\rangle $ and $\left\langle E_{l}^{2}\right\rangle $, $%
l=1,\ldots ,D-1$, in Minkowski bulk. For $D=3$ and $D=5$, the leading term (%
\ref{E2close}), coincides with the exact expressions (\ref{E2D3}) and (\ref%
{E2bD5}). By taking into account that $\zeta _{p}(x)>\zeta (p)$ for $0<x<1$,
from (\ref{E2close}) we conclude that at small separations the mean electric
field squared is positive.

When the proper separation between the plates is large, $L/\eta \gg 1$, the
effects from the curvature of the background spacetime are essential.
Assuming that $z^{D},L-z^{D}\gg \eta $ and by using the asymptotic formulas (%
\ref{Gdas}), we get
\begin{equation}
\left\langle E^{2}\right\rangle _{\mathrm{b}}\approx \frac{(D-1)\Gamma
\left( D/2-2\right) }{8\pi ^{D/2}\alpha ^{D+1}\left( L/\eta \right) ^{6}}%
\left[ (D-6)\zeta (6)-\left( D/2-4\right) \zeta _{6}(z^{D}/L)\right] ,
\label{E2far1}
\end{equation}%
for $D>4$ and
\begin{equation}
\left\langle E^{2}\right\rangle _{\mathrm{b}}\approx -\frac{2^{3-2D}\Gamma
(D)\Gamma (2-D/2)}{\pi ^{D/2}\alpha ^{D+1}\left( L/\eta \right) ^{2(D-1)}}%
\left[ 2\left( D-2\right) \zeta (2(D-1))-D\zeta _{2(D-1)}(z^{D}/L)\right] ,
\label{E2far2}
\end{equation}%
for $D<4$. In the case $D=4$ the asymptotic has the form%
\begin{equation}
\left\langle E^{2}\right\rangle _{\mathrm{b}}\approx -\frac{3\alpha ^{-5}}{%
\pi ^{2}\left( L/\eta \right) ^{6}}\left[ \sum_{n=1}^{+\infty }\frac{\ln
\left( nL/\eta \right) }{n^{6}}-\sum_{n=-\infty }^{+\infty }\frac{\ln \left(
\left\vert z^{D}-nL\right\vert /\eta \right) }{\left( z^{D}/L-n\right) ^{6}}%
\right] .  \label{E2far3}
\end{equation}%
Again, for $D=3$ and $D=5$, the expressions (\ref{E2far2}) and (\ref{E2far1}%
) coincide with the exact results (\ref{E2D3}) and (\ref{E2bD5}),
respectively. With dependence of $D$ and $z^{D}/L$, the boundary-induced VEV
$\left\langle E^{2}\right\rangle _{\mathrm{b}}$ can be either positive or
negative.

\subsection{Energy-momentum tensor}

Another important characteristic of the vacuum state is the VEV of the
energy-momentum tensor. It describes the local structure of the vacuum state
and plays an important role in modeling the self-consistent dynamic
involving the gravitational field.

Having the two-point function for the electromagnetic field tensor we can
evaluate the VEV of the energy-momentum tensor by making use of the formula%
\begin{equation}
\left\langle T_{\mu \nu }\right\rangle =\frac{1}{4\pi }g^{\beta \rho
}\lim_{x^{\prime }\rightarrow x}\left[ -W_{\mu \beta ,\nu \rho }(x,x^{\prime
})+\frac{1}{4}g_{\mu \nu }g^{\sigma s}W_{\sigma \beta ,s\rho }(x,x^{\prime })%
\right] .  \label{EMT}
\end{equation}%
Of course, the expression in the right-hand side is divergent and some
renormalization procedure is needed. The important thing here is that we
have already decomposed the two-point functions. This allows us to have a
similar decomposition of the vacuum energy-momentum tensor:%
\begin{equation}
\left\langle T_{\mu }^{\nu }\right\rangle =\left\langle T_{\mu }^{\nu
}\right\rangle _{0}+\left\langle T_{\mu }^{\nu }\right\rangle _{\mathrm{b}},
\label{DecEMT}
\end{equation}%
with the boundary-free, $\left\langle T_{\mu }^{\nu }\right\rangle _{0}$,
and boundary-induced, $\left\langle T_{\mu }^{\nu }\right\rangle _{\mathrm{b}%
}$, parts. For points outside the plates the local geometry is not changed
by the presence of the plates and the divergences in $\left\langle T_{\mu
}^{\nu }\right\rangle $ and $\left\langle T_{\mu }^{\nu }\right\rangle _{0}$
are the same. Hence, with the decomposition (\ref{DecEMT}), for the points
outside the plate the renormalization of $\left\langle T_{\mu }^{\nu
}\right\rangle $ is reduced to the one for boundary-free dS spacetime.
Because of the maximal symmetry of de Sitter spacetime and the Bunch-Davies
vacuum state, the corresponding VEV of the energy-momentum tensor is
proportional to the metric tensor: $\left\langle T_{\mu }^{\nu
}\right\rangle _{0}=\mathrm{const}\cdot \delta _{\mu }^{\nu }$.

Our main interest here is the boundary-induced part in the VEV of the
energy-momentum tensor. It is obtained by using the formula (\ref{EMT}) with
the function $W_{\mu \beta ,\nu \rho }^{(b)}(x,x^{\prime })$. In the region
between the plates, for the diagonal components of vacuum energy-momentum
tensor one finds (no summation over $\mu $)%
\begin{equation}
\left\langle T_{\mu }^{\mu }\right\rangle _{\mathrm{b}}=\frac{D-1}{16\pi
\alpha ^{D+1}B_{D}}\left[ 2\sum_{n=1}^{+\infty }U^{\left( \mu \right)
}\left( u_{n}\right) +\sum_{n=-\infty }^{+\infty }V^{\left( \mu \right)
}\left( v_{n}\right) \right] .  \label{Tmu}
\end{equation}%
Here we have introduced new functions
\begin{eqnarray}
U^{\left( 0\right) }\left( u\right)  &=&\left[ 2\left( u-1\right) \partial
_{u}+D\right] F_{D}^{(+)}(u),  \notag \\
U^{\left( l\right) }\left( u\right)  &=&\left[ 2\frac{D-3}{D-1}\left(
u-1\right) \partial _{u}+D-2\right] F_{D}^{(-)}(u)-2F_{D}(u),  \label{g} \\
U^{\left( D\right) }\left( u\right)  &=&-\left[ 2\left( u-1\right) \partial
_{u}+D-2\right] F_{D}^{(+)}(u)+2(D-3)F_{D}(u),  \notag
\end{eqnarray}%
and%
\begin{eqnarray}
V^{\left( 0\right) }\left( u\right)  &=&\left[ 2\left( u-1\right) \partial
_{u}+D-2\right] F_{D}^{(-)}(u)-2\left( D-3\right) F_{D}(u),  \notag \\
V^{\left( l\right) }\left( u\right)  &=&2\frac{D-3}{D-1}\left[ \left(
u-1\right) \partial _{u}F_{D}^{(+)}(y)+2F_{D}(u)\right] -(D-4)F_{D}^{(-)}(u),
\label{f} \\
V^{\left( D\right) }\left( u\right)  &=&-\left[ 2\left( u-1\right) \partial
_{u}+D\right] F_{D}^{(-)}(u),  \notag
\end{eqnarray}%
with $l=1,2,....,D-1$ and
\begin{equation}
F_{D}^{(\pm )}(u)=F_{D}(u)\pm G_{D}(u).  \label{FDpm}
\end{equation}%
The expression (\ref{Tmu}) is symmetric with respect to the plane $z^{D}=L/2$%
. In (\ref{Tmu}), the contributions of the terms $V^{\left( \mu \right)
}\left( v_{0}\right) $ and $V^{\left( \mu \right) }\left( v_{+1}\right) $
correspond to the geometries with single plates at $z^{D}=0$ and $z^{D}=L$,
respectively, when the second plate is absent. For $D\geqslant 3$, the
divergences on the plates come from these terms.

The problem under consideration is not symmetric under the translations
along the time coordinate and along the $z^{D}$ coordinate. As a result of
this, the vacuum energy-momentum tensor has nonzero off-diagonal component
\begin{equation}
\left\langle T_{0}^{D}\right\rangle _{\mathrm{b}}=-\frac{(D-1)L/\eta }{4\pi
B_{D}\alpha ^{D+1}}\sum_{n=-\infty }^{+\infty }(z^{D}/L-n)\left[ \left(
v_{n}-1\right) \partial _{v_{n}}+2\right] F_{D}(v_{n}),  \label{T0Dn}
\end{equation}%
which describes the energy flux along the direction normal to the plates.
The off-diagonal component is antisymmetric with respect to the plane $%
z^{D}=L/2$ and, hence, it vanishes at $z^{D}=L/2$. The $\ n=0$ term in the
right-hand side of (\ref{T0Dn})\ gives the energy flux induced by the plate
at $z^{D}=0$ when the right plate is absent. Similarly, the $n=1$ term
presents the energy flux in the geometry of a single plate $z^{D}=L$. The
boundary-induced VEVs (\ref{Tmu}) and (\ref{T0Dn}) depend on $L$, $z^{D}$,
and $\eta $ through the combinations $L/\eta $ and $z^{D}/\eta $. As we have
noticed before, they are the proper separation between the plates and the
proper distance from the left plate measured in units of $\alpha $.

For $D\geqslant 4$, for points near the plates the VEVs are dominated by
single plate parts. In particular, near the left plate, to the leading order
one has \cite{Saha14} (no summation over $l=0,\ldots ,D-1$)%
\begin{equation}
\langle T_{l}^{l}\rangle _{\mathrm{b}}\approx -\frac{\eta }{z^{D}}\langle
T_{0}^{D}\rangle _{\mathrm{b}}\approx \frac{D-1}{(z^{D}/\eta )^{2}}\langle
T_{D}^{D}\rangle _{\mathrm{b}}\approx -\frac{\left( D-3\right) (D-1)\Gamma
((D+1)/2)}{2(4\pi )^{(D+1)/2}\alpha ^{D+1}(z^{D}/\eta )^{D+1}}.
\label{Tllnear}
\end{equation}%
The asymptotic expressions near the right plate are obtained by the
replacement $z^{D}\rightarrow L-z^{D}$. As is seen, near the plates the
diagonal components of the boundary-induced vacuum energy-momentum tensor
are negative and the energy flux is directed from the plates.

By using $F_{3}(z)=2/(1-z)^{2}$ and the expression for $G_{3}\left( y\right)
$ given above, we can see that for $D=3$ one has%
\begin{equation}
\left\langle T_{\mu }^{\nu }\right\rangle _{\mathrm{b}}=-\frac{\pi ^{2}}{%
720\left( \alpha L/\eta \right) ^{4}}\mathrm{diag}(1,1,1,-3).  \label{EMTD3}
\end{equation}%
This expression coincides with the standard result for the plates in
Minkowski bulk \cite{Brow69} with the separation $\alpha L/\eta $. Again,
this property is a consequence of the conformal invariance of the
electromagnetic field in $D=3$. Relatively simple expressions are also
obtained for $D=5$. In this case for the diagonal components we get (no
summation over $\mu $)%
\begin{equation}
\left\langle T_{\mu }^{\mu }\right\rangle _{\mathrm{b}}=\frac{\left( \alpha
L/\eta \right) ^{-6}}{32\pi ^{3}}\sum_{m=2,3}\frac{(-1)^{m}}{(L/\eta
)^{2(m-3)}}\left[ 2U^{\left( \mu ,m\right) }\zeta (2m)+V^{\left( \mu
,m\right) }\zeta _{2m}(z^{D}/L)\right] ,  \label{TmuD5}
\end{equation}%
with the coefficients%
\begin{eqnarray}
&&(U^{\left( 0,2\right) },U^{\left( 1,2\right) },U^{\left( 5,2\right)
})=(1,-1,5),\;(U^{\left( 0,3\right) },U^{\left( 1,3\right) },U^{\left(
5,3\right) })=(2,2,-10),  \notag \\
&&(V^{\left( 0,2\right) },V^{\left( 1,2\right) },V^{\left( 5,2\right)
})=(-5,-1,-1),\;(V^{\left( 0,3\right) },V^{\left( 1,3\right) },V^{\left(
5,3\right) })=(4,4,0).  \label{f02}
\end{eqnarray}%
For $D=5$ the expression of the off-diagonal component takes the form%
\begin{equation}
\left\langle T_{0}^{D}\right\rangle _{\mathrm{b}}=\frac{\zeta
(5,z^{D}/L)-\zeta (5,1-z^{D}/L)}{8\pi ^{3}\alpha \left( \alpha L/\eta
\right) ^{5}}.  \label{T0DD5}
\end{equation}

Now we turn to the investigation of the asymptotic behavior for the
components of VEV of the energy-momentum tensor. First, we consider the
limit when the proper separation between the plates is small compared with
the dS curvature radius, \ $L/\eta \ll 1$. In this limit the gravitational
effects in the boundary-induced parts of the diagonal components are
subdominant and to the leading order one finds (no summation over $\mu $)%
\begin{equation}
\left\langle T_{\mu }^{\mu }\right\rangle _{\mathrm{b}}\approx -\frac{\left(
D-1\right) \Gamma \left( (D+1)/2\right) }{2\left( 4\pi \right)
^{(D+1)/2}\left( \alpha L/\eta \right) ^{D+1}}\left[ 2\zeta (D+1)+(D-3)\zeta
_{D+1}(z^{D}/L)\right] ,  \label{Tmufar}
\end{equation}%
for $\mu =0,1,...,D-1$, and%
\begin{equation}
\left\langle T_{D}^{D}\right\rangle _{\mathrm{b}}\approx \frac{D\left(
D-1\right) \Gamma \left( (D+1)/2\right) }{\left( 4\pi \right)
^{(D+1)/2}\left( \alpha L/\eta \right) ^{D+1}}\zeta (D+1).  \label{TDfar}
\end{equation}%
The expressions (\ref{Tmufar}) and (\ref{TDfar}) coincide with the
corresponding results for parallel plates in Minkowski bulk with the
separation $\alpha L/\eta $. For the energy flux, to the leading order, one
gets%
\begin{equation}
\left\langle T_{0}^{D}\right\rangle \approx \frac{(D-1)(D-3)\Gamma ((D+1)/2)%
}{2\left( 4\pi \right) ^{\left( D+1\right) /2}\alpha (\alpha L/\eta )^{D}}%
\left[ \zeta (D,z^{D}/L)-\zeta (D,1-z^{D}/L)\right] .  \label{T0Dfar}
\end{equation}%
The flux is positive for $0<z^{D}<L/2$ and negative for $L/2<$ $z^{D}<L$.
Note that for $D=5$ the expression in the right-hand side of (\ref{T0Dfar})
coincides with the exact result (\ref{T0DD5}).

In the opposite limit, when the separation between the plates is large
compared to the dS curvature radius, that corresponds to $L/\eta \gg 1$, and
assuming also $z^{D},$ $L-z^{D}\gg \eta $, for $D\geqslant 4$ to the leading
order one finds (no summation over $\mu $)%
\begin{equation}
\left\langle T_{\mu }^{\mu }\right\rangle _{\mathrm{b}}\approx \frac{%
(D-1)\Gamma \left( D/2-1\right) }{2^{6}\pi ^{D/2+1}\alpha ^{D+1}(L/\eta )^{4}%
}\left[ 2U_{\mu }\zeta (4)+V_{\mu }\zeta _{4}(z^{D}/L)\right] ,
\label{EMTlarge}
\end{equation}%
with the notations%
\begin{eqnarray}
U_{0} &=&D-4,\;U_{l}=D-4-4\frac{D-3}{D-1},\;U_{D}=D,  \notag \\
\;V_{0} &=&-D,\;V_{l}=V_{D}=4-D,  \label{f0}
\end{eqnarray}%
and $l=1,\ldots ,D-1$. For the off-diagonal component, in the same limit, we
get%
\begin{equation}
\langle T_{0}^{D}\rangle _{\mathrm{b}}\approx \frac{(D-1)\Gamma \left(
D/2-1\right) }{2^{4}\pi ^{D/2+1}\alpha ^{D+1}(L/\eta )^{5}}\left[ \zeta
(5,z^{D}/L)-\zeta (5,1-z^{D}/L)\right] .  \label{T0Dlarge}
\end{equation}%
Again, the flux is positive for $0<z^{D}<L/2$ and negative for $L/2<z^{D}<L$.

\begin{figure}[tbph]
\begin{center}
\begin{tabular}{cc}
\epsfig{figure=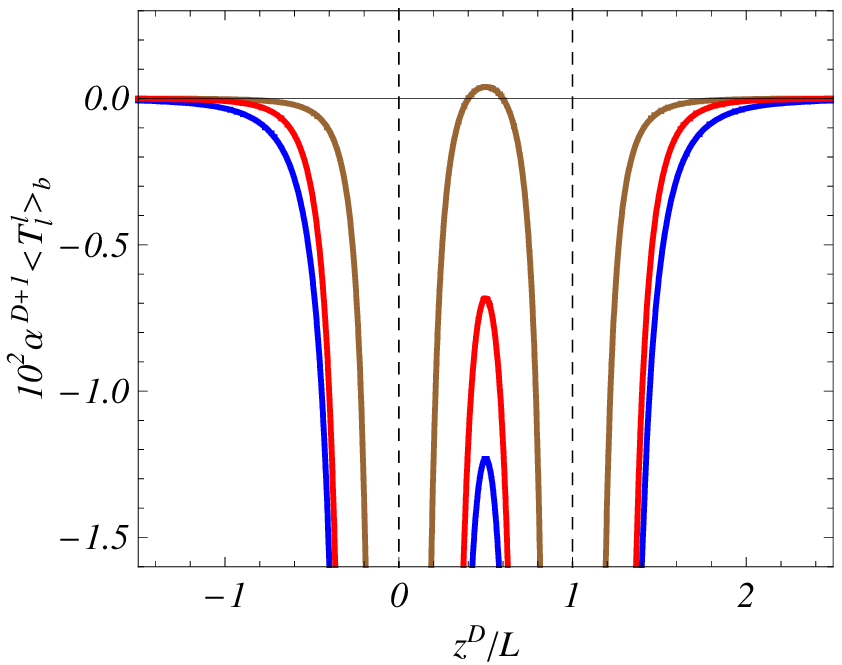,width=7.cm,height=5.5cm} & \quad %
\epsfig{figure=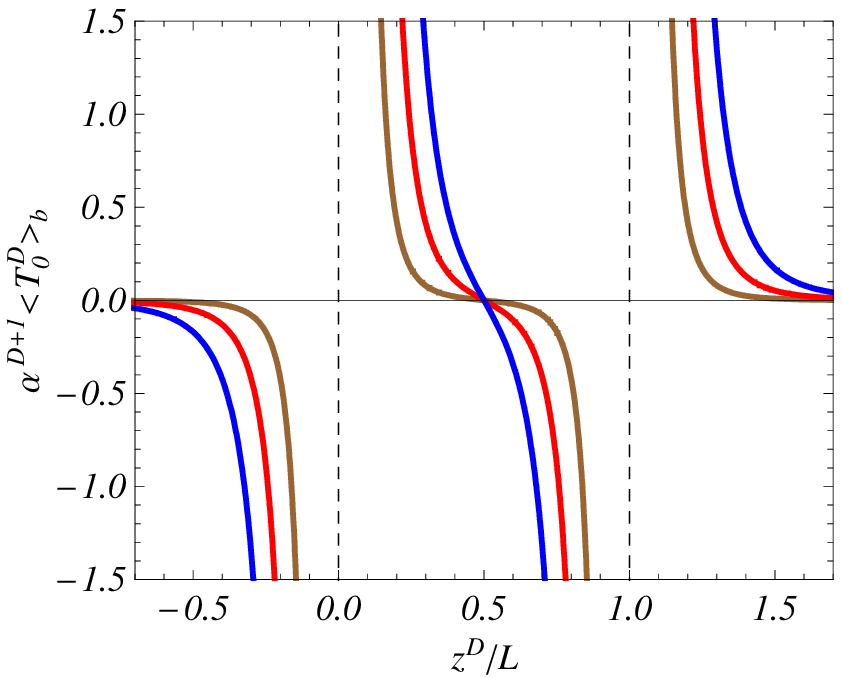,width=7.cm,height=5.5cm}%
\end{tabular}%
\end{center}
\caption{The VEVs of the diagonal (left panel) and off-diagonal components
of the energy-momentum tensor as functions of the rescaled coordinate $%
z^{D}/L$ in the geometry of two conducting plates for $D=4$. The vertical
dashed lines correspond to the locations of the plates.}
\label{fig1}
\end{figure}

In figure \ref{fig1} we have plotted the diagonal (left panel) and the
off-diagonal (right panel) components of the vacuum energy-momentum tensor
as functions of $z^{D}/L$ for $D=4$ dS spacetime. For the left panel we have
taken the separation between the plates corresponding to $L/\eta =2$ and the
lower, middle, upper curves correspond to the components $\left\langle
T_{0}^{0}\right\rangle _{\mathrm{b}}$, $\left\langle T_{1}^{1}\right\rangle
_{\mathrm{b}}$, $\left\langle T_{4}^{4}\right\rangle _{\mathrm{b}}$,
respectively. On the left panel the energy flux is plotted for $L/\eta
=0.75,1,1.5$ (the curves from the right to the left in the regions $%
0<z^{D}/L<1/2$, $z^{D}/L>1$, and the curves from the left to the right in
the regions $z^{D}/L<0$, $1/2<z^{D}/L<1$).

\section{Casimir forces}

\label{sec:Force}

From the symmetry of the problem under consideration it follows that the
Casimir forces acting on the left and right plates are equal in magnitude.
We shall consider the force acting per unit surface of the left plate (the
vacuum effective pressure on the plate). Having the normal stress, the
effective pressure $P$ is evaluated as $P=-\left\langle
T_{D}^{D}\right\rangle _{z^{D}=0}$. In the discussion above we have
decomposed the normal stress into the single plate and second plate-induced
contributions. Consequently, the Casimir force is decomposed into the
self-action and interaction parts. The self-action force on the right-hand
side of the plate at $z^{D}=0$ is determined by the $n=0$ term in the
expression (\ref{Tmu}) for the component with $\mu =D$. Because of the
surface divergences in the boundary-induced VEVs this force is divergent.
However, by the symmetry, the self-forces acting on a single plate from the
left- and right-hand sides cancel each other and the net self-action force
vanishes (note that this would not be the case for curved boundaries).

The interaction part of the vacuum pressure, denoted here as $P^{\left(
\mathrm{int}\right) }$, is induced by the presence of the second plate. It
is directly obtained from the expression (\ref{Tmu}) for $\left\langle
T_{D}^{D}\right\rangle $ omitting the term $n=0$:
\begin{equation}
P^{\left( \mathrm{int}\right) }=\frac{(D-1)\alpha ^{-D-1}}{(4\pi
)^{(D+1)/2}\Gamma \left( \left( D+3\right) /2\right) }\sum_{n=1}^{\infty
}\left\{ 2\left[ \left( u_{n}-1\right) \partial _{u_{n}}+1\right]
F_{D}\left( u_{n}\right) -G_{D}(u_{n})\right\} ,  \label{Pint}
\end{equation}%
where $u_{n}$ is defined by (\ref{vn}). Relatively simple expressions are
obtained for odd values of the spatial dimension. In particular, in the
cases $D=3$ and $D=5$ one finds%
\begin{eqnarray}
P^{\left( \mathrm{int}\right) } &=&-\frac{\pi ^{2}}{240(\alpha L/\eta )^{4}}%
,\;D=3  \notag \\
P^{\left( \mathrm{int}\right) } &=&-\frac{\pi \alpha ^{-6}}{360(L/\eta )^{4}}%
\left[ 1+\frac{5\pi ^{2}}{21(L/\eta )^{2}}\right] ,\;D=5.  \label{PintD5}
\end{eqnarray}

For small separations between the plates, that correspond to\ $L/\eta \ll 1$%
, from (\ref{Pint}) one has the asymptotic expression%
\begin{equation}
P^{\left( \mathrm{int}\right) }\approx -\frac{D(D-1)\Gamma \left(
(D+1)/2\right) }{\left( 4\pi \right) ^{\left( D+1\right) /2}\left( \alpha
L/\eta \right) ^{D+1}}\zeta \left( D+1\right) .  \label{Pintsmall}
\end{equation}%
This leading term coincides with the expression of the force in Minkowski
bulk with the separation between the plates equal to $\alpha L/\eta $
(proper distance between the plates in dS spacetime). At large separations
between plates one has \ $L/\eta \gg 1$ and for $D\geqslant 4$ to the
leading order we get
\begin{equation}
P^{\left( \mathrm{int}\right) }\approx -\frac{\left( D-1\right) \Gamma
(D/2-1)}{720\pi ^{D/2-3}\alpha ^{D+1}\left( L/\eta \right) ^{4}}.
\label{Pintlarge}
\end{equation}%
In both cases of (\ref{Pintsmall}) and (\ref{Pintlarge}) we have $P^{\left(
\mathrm{int}\right) }<0$ and the corresponding forces are attractive. It is
of interest to note that at large separations the force decays as the
inverse forth power of the proper separation for all values of $D\geqslant 3$%
. As it has been shown in \cite{Saha09}, for a massive scalar field on dS
background, depending on the values of the curvature coupling parameter and
the mass of the field, two different regimes are realized, which exhibit
monotonic and oscillatory decay of the Casimir forces at large separations.
In the physical problem under consideration the quantum field is massless
and the forces decay monotonically.

By taking into account that $\eta =\alpha e^{-t/\alpha }$, we see that the
formulae (\ref{Pintsmall}) and (\ref{Pintlarge}) describe the behavior of
the Casimir forces at early and late stages of the cosmological expansion
for a fixed value of the ratio $L/\alpha $. At early stages of the
expansion, $t\rightarrow -\infty $, the Casimir force behaves as $P^{\left(
\mathrm{int}\right) }\propto e^{-(D+1)t/\alpha }$. At late stages, $%
t\rightarrow +\infty $, the force decays as $P^{\left( \mathrm{int}\right)
}\propto e^{-4t/\alpha }$ for all values of the spatial dimension $%
D\geqslant 3$.

For the study of physical effects induced by boundaries in dS spacetime we
have considered planar coordinates having flat spatial sections. These
coordinates only cover half of dS spacetime. But this does not cause any
problem from the cosmological point of view. From this perspective the
background geometry under consideration is just a special case of the larger
class of spatially flat Friedmann-Robertson-Walker-type models, which are
relevant to inflationary cosmology. Moreover, the fact that planar
coordinates cover a part of dS spacetime does not cause problems in the
quantization of fields because $t=\mathrm{const}$ defines a Cauchy surface
and the information from any part of the dS manifold enters the submanifold
covered by these coordinates as an initial condition (for a recent
discussion see \cite{Degu13}). The full dS spacetime is covered by global
coordinates with spherical spatial sections. An important point is that the
Bunch-Davies state defines the vacuum states in both planar and global
coordinates. As a result, in global coordinates, the two-point functions and
the VEVs for the region between the boundaries are obtained from the
corresponding expressions in planar coordinates by using the standard
transformation formulae for the components of tensors. Note that in the
problem under consideration the planar coordinates are the most appropriate
ones because the boundaries are coordinate surfaces. The latter is not the
case in global coordinates and the equation of the boundaries in these
coordinates is complicated.

\section{Conclusion}

\label{sec:Conc}

We have considered the Casimir effect for the electromagnetic field in the
geometry of two planar conducting boundaries on background of $(D+1)$%
-dimensional dS spacetime. For the Bunch-Davies vacuum state the two-point
function for the electromagnetic field tensor is evaluated in the region
between the plates. This function is expressed in terms of the boundary-free
function by the formula (\ref{WF}) and the latter is given by (\ref{FF0}).
We consider a free field theory and all the information on the properties of
the vacuum state is encoded in the two-point functions. As important local
characteristics of the vacuum state, we evaluate the mean field squared and
the VEV of the energy-momentum tensor. These VEVs are decomposed into the
boundary-free and plate-induced parts. For points away from the boundaries,
the renormalization is required for the boundary-free parts only. For $D=3$
the electromagnetic field is conformally invariant and the plate-induced
VEVs are expressed in terms of the corresponding quantities for conducting
plates in Minkowski spacetime by standard conformal relations (see (\ref%
{E2D3}) and (\ref{EMTD3})).

In the region between the plates, the boundary-induced part in the VEV of
the electric field squared is given by the expression (\ref{E2}) where the
function $G_{D}(z)$ is expressed in terms of the hypergeometric function by (%
\ref{FD}). For the proper separation between the plates smaller than the dS
curvature radius, the leading term in the electric field squared is given by
(\ref{E2close}) and this term coincides with the corresponding VEV for the
plates in Minkowski spacetime with the separation $\alpha L/\eta $. In this
limit the mean electric field squared is positive. In the opposite limit of
large separations, assuming that $z^{D},L-z^{D}\gg \eta $, the leading terms
in the asymptotic expansion are given by (\ref{E2far1}) for $D>4$ and by (%
\ref{E2far2}) for $D<4$. For $D=4$ one has the asymptotic (\ref{E2far3}). At
large separations, with dependence of $D$ and $z^{D}/L$, the
boundary-induced VEV in the electric field squared can be either positive or
negative. In the Minkowski bulk this VEV is always positive.

For the boundary-induced part in the VEV of the diagonal components
of the energy-momentum tensor we have derived the expression
(\ref{Tmu}) with the functions (\ref{g}) and (\ref{f}). In addition,
for $D\neq 3$, the vacuum energy-momentum tensor has a nonzero
off-diagonal component (\ref{T0Dn}) that describes the energy flux
along the direction normal to the plates. The off-diagonal component
is antisymmetric with respect to the plane $z^{D}=L/2$ and, hence,
it vanishes at $z^{D}=L/2$. In the region between the plates, the
flux is positive for $z^{D}<L/2$ and negative for $z^{D}>L/2$. At
small separations, the VEVs of the diagonal components of the
energy-momentum tensor, to the leading order, coincide with the
corresponding results for the plates in Minkowski spacetime with the
separation $\alpha L/\eta $, whereas the leading term in the
asymptotic expansion of the off-diagonal component is given by
(\ref{T0Dfar}). In the limit when the separation between the plates
is large compared to the dS curvature radius and assuming that
$z^{D},$ $L-z^{D}\gg \eta $, for $D\geqslant 4$ the terms in the
asymptotic expansions are given by the expressions (\ref{EMTlarge}) and (\ref%
{T0Dlarge}).

The interaction part of the Casimir pressure on the plates is given by the
expression (\ref{Pint}) and the corresponding forces are attractive. At
small separations, to the leading order we obtain the Casimir force for the
geometry of two parallel plates in Minkowski spacetime with the separation $%
\alpha L/\eta $. For large separations and for $D\geqslant 4$, the leading
term in the corresponding asymptotic expansion is given by (\ref{Pintlarge}%
). At large separations the force decays as the inverse forth power of the
proper separation for all values of $D\geqslant 3$. This is in contrast with
the case of the plates in Minkowski bulk where the force decays inversely
proportional to the $(D+1)$th power of the separation. Hence, for $%
D\geqslant 4$, in dS spacetime the decay of the Casimir forces at large
separations between the plates is weaker than that for the plates in
Minkowski bulk.

In the discussion above we have imposed the boundary condition which is a
generalization of the perfectly conducting boundary condition for
higher-dimensional models. Instead, we could impose the boundary condition $%
n^{\mu }F_{\mu \nu }=0$ that is used in MIT bag models of hadrons
for the confinement of the color. The corresponding modes are
obtained from (\ref{Modes1}) with the replacement $\sin (
k_{D}z^{D})\rightarrow \cos ( k_{D}z^{D})$ and from (\ref{Modes2})
with the replacement $\cos ( k_{D}z^{D})\rightarrow \sin (
k_{D}z^{D})$. The further evaluation of the two-point functions and
VEVs is similar to that we have described above. The corresponding
results for the both types of boundary conditions may have
applications in braneworlds on dS bulk with reflecting branes.
Bearing in mind further applications in Klauza-Klein-type models
with compact extra dimensions, it would be interesting to generalize
the expressions for the Casimir densities in the case of a locally
dS spacetime a part of spatial coordinates of which are
compactified. The comparison of the Casimir force with the
corresponding expression for the geometry in the absence of compact
dimensions would give constraints on the geometry of a compact
subspace (for a discussion in AdS bulk see \cite{Teo10}).

\section*{Acknowledgments}

A. A. S. and H. A. N. were supported by the State Committee of Science
Ministry of Education and Science RA, within the frame of Grant No. SCS
13-1C040.

\end{document}